\documentclass[aps,pre,reprint,superscriptaddress,nobibnotes,floatfix]{revtex4-2}

\usepackage[T1]{fontenc}
\usepackage[utf8]{inputenc}
\usepackage[english]{babel}
\usepackage{amsmath,amssymb,bm}
\usepackage{graphicx}
\usepackage{xcolor}
\usepackage{hyperref}
\hypersetup{colorlinks=true,citecolor=blue,linkcolor=blue,urlcolor=blue}

\newcommand{\taueq}{\tau_{\mathrm{eq}}}
\newcommand{\tauB}{\tau_{\mathrm{B}}}
\newcommand{\tauT}{\tau_{\mathrm{trap}}}

\begin{document}

\title{The Fiedler dimension of networks of networks: from fractal to small-world architectures}

\author{Lorenzo Grimaldi}
\email{lorenzo.grimaldi@cref.it}
\affiliation{`Enrico Fermi' Research Center (CREF), Via Panisperna 89A, 00184 Rome, Italy}
\affiliation{Dipartimento di Fisica dell'Universit\`a di Roma `Tor Vergata' e sezione INFN di Roma `Tor Vergata', Via della Ricerca Scientifica 1, 00133 Rome, Italy}

\author{Andrea Gabrielli}
\email{andrea.gabrielli@cref.it}
\affiliation{`Enrico Fermi' Research Center (CREF), Via Panisperna 89A, 00184 Rome, Italy}
\affiliation{Dipartimento di Ingegneria Civile, Informatica e delle Tecnologie Aeronautiche, Universit\`a degli Studi `Roma Tre', Via Vito Volterra 62, 00146 Rome, Italy}

\author{Pablo Villegas}
\email{pablo.villegas@cref.it}
\affiliation{`Enrico Fermi' Research Center (CREF), Via Panisperna 89A, 00184 Rome, Italy}
\affiliation{Instituto Carlos I de F\'isica Te\'orica y Computacional, Universidad de Granada, E-18071 Granada, Spain}

\begin{abstract}
Is the relaxation of a modular network dictated by its modules or by the network that connects them? The answer fixes the time scales of diffusion, consensus, and synchronization, all set by the Fiedler eigenvalue. We show that, for bundled networks, the two levels act one after the other: a random walker must first escape from its module and then spread over the network that connects them. The equilibration time is the sum of these two times, the time needed to reach the base from inside a fiber and the relaxation time of the base, slowed down by the mass of the fibers. The consequences are unexpected. However large the modules are, a small-world core always imposes its own Fiedler dimension; the modules survive only as a logarithmic correction, which slows down equilibration and hides the true exponent up to sizes far beyond any real network. Effective dimensions measured on finite modular systems can therefore be systematically biased. We analytically work out all combinations of finite-dimensional and small-world bases and fibers, and compare them with numerically exact spectra of representatives of each category.
\end{abstract}

\maketitle

\section{Introduction}

In a system relaxing toward equilibrium, the longest time scale is set by the slowest mode. On a network, random walks, diffusion, and consensus dynamics are generated by the graph Laplacian, and their slowest mode is fixed by its smallest nonzero eigenvalue, the Fiedler eigenvalue $\lambda_g$~\cite{Fiedler_eig}. The equilibration time is $\lambda_g^{-1}$; the relaxation toward synchronization of coupled oscillators is controlled by $\lambda_g$~\cite{Pecora1998,Barahona2002}, and a small $\lambda_g$ signals a bottleneck that splits the network into weakly coupled parts. Real networks are full of such bottlenecks because they are modular. Many are even assembled by attaching pre-existing modules to a common core~\cite{Jacob1977}: genomes acquire genes horizontally~\cite{Crisp2015,Pace2008}, neural circuits reuse conserved blocks~\cite{Grillner2016}, and software grows by recombining packages~\cite{Villegas2020,Fortuna2011}. Modularity separates fast relaxation inside the modules from slow equilibration among them~\cite{Arenas2006}, and coupling two networks can change $\lambda_g$ abruptly~\cite{RadicchiArenas2013,Gomez2013}. Which of the two levels sets the slowest time scale of a large modular network, the modules or the way they are wired together, and how this time grows with the number of nodes, remain open questions.

The way relaxation slows down as the network grows is captured by two dimensions. The spectral dimension $d_s$ describes the density of slow modes~\cite{AlexanderOrbach,Hattori1987,Cassi1992,Burioni1996,RW_review}; the Fiedler dimension $d_g$, defined by $\lambda_g\sim N^{-2/d_g}$, describes the slowest one. Effective dimensions of this kind are increasingly used to characterize real networks, including brain networks~\cite{Peach2022,Poggialini2025}. In homogeneous structures $d_g=d_s$. In previous work~\cite{Grimaldi_2026}, we showed that this identity fails in composite networks, so that networks with the same thermodynamic exponents can equilibrate on parametrically different time scales. That analysis used bundled networks, in which a copy of a fiber graph is attached to every node of a base graph~\cite{Cassi1996_Bundled,spectral_dim_bund}, and assumed that both have finite fractal dimensions. Many real modular networks are instead small-world: their diameter grows as $\ln N$~\cite{WattsStrogatz1998,DorogovtsevRMP2008}, or even more slowly~\cite{Cohen2003}. Deterministic hierarchical models, such as the pseudofractal web~\cite{DGM2002} and the $(1,v)$-flowers~\cite{Rozenfeld_2007}, are transfractals, whose mass grows exponentially with the diameter. For such architectures the finite-dimensional theory does not apply, and simply sending the fractal dimension of the base to infinity in it is not justified a priori. It is not even clear whether the modules still matter once they hang on a small-world core, nor how reliable the effective dimensions measured on finite modular networks are.

Here we answer these questions with three results. First, the equilibration time of a bundle is the sum of two different times with a transparent meaning: the time needed to reach the base graph from inside a fiber, and the relaxation time of the base, slowed down by the mass of the fibers. The two processes act in series, since the walker cannot feel the base before reaching it, so their times add like resistances in series and the slower one controls equilibration. Second, this series law turns into a composition rule for $d_g$ that depends only on which component, base or fibers, dominates the growth of the total mass. Third, a small-world base always wins: it imposes its own Fiedler dimension, while the fibers survive only as a logarithmic correction that slows down equilibration and biases finite-size estimates of $d_g$ over many decades. We check these predictions in all four combinations of finite-dimensional and small-world bases and fibers, with exact spectra of different scale-invariant networks.

\section{Diffusion scales in composite networks}
\label{sec:Diffusion scales in composite networks}

We follow the notation of Ref.~\cite{Grimaldi_2026}. For a connected undirected graph with Laplacian $\hat L=\hat D-\hat A$ and eigenvalues $0=\lambda_0<\lambda_1\le\dots\le\lambda_{N-1}$, diffusion scales are probed by the Laplacian density matrix $\hat\rho(\tau)=e^{-\tau\hat L}/Z(\tau)$~\cite{Domenico2016,LRG,Modularity}. The associated entropy $S(\tau)$ defines the diffusion-scale heat capacity~\cite{InfoCore,LRG}
\begin{equation}
    C(\tau)\equiv-\frac{\mathrm{d}S}{\mathrm{d}\ln\tau}=-\tau^2\frac{\mathrm{d}\langle\lambda\rangle_\tau}{\mathrm{d}\tau}.
    \label{eq:C}
\end{equation}
If the spectral density behaves as $\rho(\lambda)\sim\lambda^{d_s/2-1}$ at small $\lambda$, $C(\tau)$ displays a plateau at $d_s/2$~\cite{Poggialini2025,Villegas2025}. At finite size, $C(\tau)$ vanishes beyond the equilibration time $\taueq\sim\lambda_g^{-1}$, set by the Fiedler eigenvalue $\lambda_g\equiv\lambda_1$. Its scaling,
\begin{equation}
    \lambda_g\sim N^{-\frac{2}{d_g}},
    \label{eq:dg}
\end{equation}
defines the Fiedler dimension $d_g$~\cite{Fiedler_eig}. In homogeneous structures $d_g=d_s$~\cite{spectral_fiedler_instability,Villegas2025}.

\begin{figure*}[t]
\centering
\includegraphics[width=\textwidth]{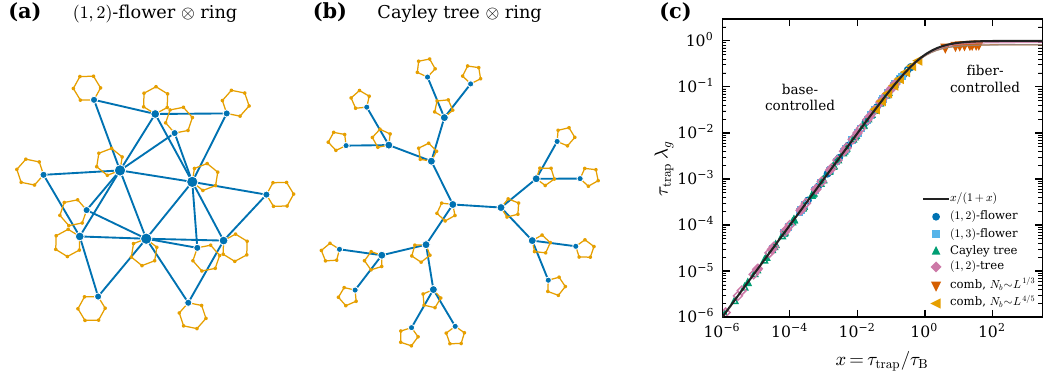}
\caption{(a) Pseudofractal $(1,2)$-flower~\cite{DGM2002,Rozenfeld_2007} and (b) Cayley tree of coordination $3$, each dressed with ring fibers. Node size grows with degree. Both bases are small-world. (c) Series law for the equilibration time, Eq.~\eqref{eq:series}: $\tauT\lambda_g$ versus $x=\tauT/\tauB$. Symbols: exact Fiedler eigenvalues for $(1,2)$- and $(1,3)$-flower, Cayley-tree, and $(1,2)$-tree bases with ring (filled), square-lattice, and cubic-lattice (open) fibers, together with the Dirac combs of Ref.~\cite{Grimaldi_2026}, whose base grows as $L^{1/3}$ or $L^{4/5}$ with fiber length $L$. Thin lines: ring, chain, and random-tree fibers with a continuously varying base eigenvalue $\ell^b_1$. Black line: $x/(1+x)$. In the fiber-controlled regime, rings and chains saturate at $\tauT\lambda_g\simeq\pi^2/12$.}
\label{fig:bundles}
\end{figure*}

A bundled network is obtained by attaching a copy of a fiber graph (Laplacian $\hat L^f$, $N_f$ nodes, root $i_2=0$) to each of the $N_b$ nodes of a base graph (Laplacian $\hat L^b$). Fig.s~\ref{fig:bundles}(a) and \ref{fig:bundles}(b) show two examples. Diagonalizing $\hat L^b$, with eigenvalues $0=\ell^b_0<\ell^b_1\le\dots$, the eigenvalue problem of the whole network splits into $N_b$ independent fiber problems~\cite{spectral_dim_bund,Boson_comb2,Grimaldi_2026},
\begin{equation}
    \sum_{j_2}L^f_{i_2,j_2}\tilde\psi_{n_1,j_2}+\ell^b_{n_1}\delta_{i_2,0}\tilde\psi_{n_1,i_2}=\lambda\,\tilde\psi_{n_1,i_2}.
    \label{eq:impurity}
\end{equation}
Each base eigenvalue thus acts as a local impurity at the root of a fiber. Since the impurity occupies a single site, the bulk density of states is that of the fiber, and $d_s=d_{s,f}$~\cite{spectral_dim_bund,RW_review}. The low-energy states, instead, originate from the smallest $\ell^b_{n_1}$. To first order, the lowest level of Eq.~\eqref{eq:impurity} is $\lambda_{n_1}\simeq\ell^b_{n_1}/N_f$~\cite{Grimaldi_2026}. With $\ell^b_1\sim N_b^{-2/d_{g,b}}$, $N_b\sim L^{d_{f,b}}$, and $N_f\sim L^{d_{f,f}}$, this gives~\cite{Grimaldi_2026}
\begin{equation}
    d_g=2\,\frac{d_{f,f}+d_{f,b}}{d_{f,f}+2\frac{d_{f,b}}{d_{g,b}}} .
    \label{eq:old}
\end{equation}
While Eq.~\eqref{eq:old} predicts the Fiedler dimension of bundled graphs where both the bulk and the fiber display a power-law growth, it does not directly apply to exponentially growing components. Moreover, the range of validity of this perturbative result has not yet been established. In \cite{Grimaldi_2026} we argued that, under specific circumstances regarding the growth of the fiber, the Fiedler eigenvalue of the whole network might be regulated by the fiber only, rather than being affected by the bulk as well. In the following, we will discuss this issue from the point of view of relaxation times.

\section{Relaxation times add in series}
\label{sec:series}
The Fiedler eigenvalue of the bundle is given by the lowest level of Eq.~\eqref{eq:impurity} with $n_1=1$. Indeed, the sector $n_1=0$ carries no impurity, since $\ell^b_0=0$: its levels are those of the free fiber, with eigenvectors uniform over the base, so the bundle spectrum contains the zero mode and the first excited level $\lambda_{1,f}$ of the free fiber. Every other sector carries an impurity $\ell^b_{n_1}>0$ on the root, which raises its levels but acts on a single node, so it cannot push the lowest level of the sector above $\lambda_{1,f}$. Since the levels grow monotonically with the strength of the impurity, the lowest nonzero level of the whole spectrum is the lowest level of the sector with the weakest impurity, $n_1=1$ (Appendix~\ref{app:series}). The problem therefore reduces to a single fiber with one impurity of strength $\ell^b_1$ at its root, which can be solved for any fiber. The result has a simple form,
\begin{equation}
    \taueq=\lambda_g^{-1}\simeq\tauB+\tauT,
    \qquad
    \tauB=\frac{N_f}{\ell^b_1}.
    \label{eq:series}
\end{equation}
Eq.~\eqref{eq:series} becomes exact when $\lambda_g\ll\lambda_{1,f}$, and it is a lower bound for $\taueq$ in all cases. Both terms have a direct random-walk interpretation:
\begin{itemize}
    \item $\tauB$ is the relaxation time of the bare base, $1/\ell^b_1$, slowed down by the factor $N_f$. The reason is that a walker spends only a fraction $1/N_f$ of its time on the base, where the slow mode lives. This is the first-order result of Ref.~\cite{Grimaldi_2026}.
    
    \item $\tauT$ is the mean first-passage time to the root of a fiber, averaged over a uniformly distributed starting point, i.e., the global trapping time of the fiber~\cite{Condamin2007,comb_of_comb}. A walker cannot feel the base before it has reached it.
\end{itemize}

The two processes act in series, so their times add. The ratio $x=\tauT/\tauB$ is the true expansion parameter of the low-energy theory. For $x\ll1$ the gap is base-controlled, $\lambda_g\simeq\ell^b_1/N_f$. For $x\gg1$ it is fiber-controlled, and $\lambda_g$ saturates at a value of order $\tauT^{-1}$, which lies between the Dirichlet gap of the fiber and $\lambda_{1,f}$. Fig.~\ref{fig:bundles}(c) confirms the collapse $\tauT\lambda_g=x/(1+x)$ over six decades, up to $x\sim1$, for very different bases and fibers. Beyond it, only a prefactor of order one is left: the height of the saturated plateau is a property of the fiber alone, and equals $\pi^2/12$ for one-dimensional fibers, as we show in Appendix~\ref{app:series}.

The trapping time follows from the spectrum of the fiber as seen from its root. As shown in Appendix~\ref{app:series}, $\tauT=N_f(\hat L^{f+})_{00}=N_f\sum_{k\geq1}w_k/\lambda_{k,f}$, where $w_k=|\phi_k(0)|^2$ is the weight of the fiber mode $\phi_k$ on the root. We assume that the root is a typical node, i.e., that its local spectral dimension equals the bulk one, as in all cases considered here. Then $w_k\simeq1/N_f$ on average, and the sum becomes an integral over the density of states of the fiber, $\rho_f(\lambda)\sim\lambda^{d_{s,f}/2-1}$, cut off from below by the fiber gap $\lambda_{1,f}\sim N_f^{-2/d_{g,f}}$:
$\tauT\sim N_f\int_{\lambda_{1,f}}\mathrm{d}\lambda\,\lambda^{d_{s,f}/2-2}$. For $d_{s,f}=d_{g,f}$ this gives~\cite{Condamin2007,RW_review}
\begin{equation}
    \tauT\sim
    \begin{cases}
      N_f^{\frac{2}{d_{g,f}}}, & d_{g,f}<2\ \ \text{(recurrent)},\\
      N_f\ln N_f, & d_{g,f}=2,\\
      N_f, & d_{g,f}>2\ \ \text{(transient)}.
    \end{cases}
    \label{eq:trap}
\end{equation}
The three regimes reflect the recurrence properties of random walks on the fiber~\cite{RW_review}. For $d_{g,f}<2$ the walk is recurrent and explores the fiber compactly, so it reaches the root only after covering the whole fiber, which takes the relaxation time of the fiber itself, $\lambda_{1,f}^{-1}\sim N_f^{2/d_{g,f}}$. For $d_{g,f}>2$ the walk is transient and keeps visiting new sites at a finite rate, so it hits the root after visiting of order $N_f$ distinct sites, in a time of order $N_f$. At $d_{g,f}=2$ the walk is marginally recurrent, and the logarithm reflects the slow growth, $t/\ln t$, of the number of distinct sites it visits. Since $\ell^b_1\to0$, a transient fiber always gives $\tauT\ll\tauB$. A fiber-controlled gap thus requires recurrent fibers.

\section{Growth-rate composition}
\label{sec:composition}

\begin{figure*}[t]
\centering
\includegraphics[width=\textwidth]{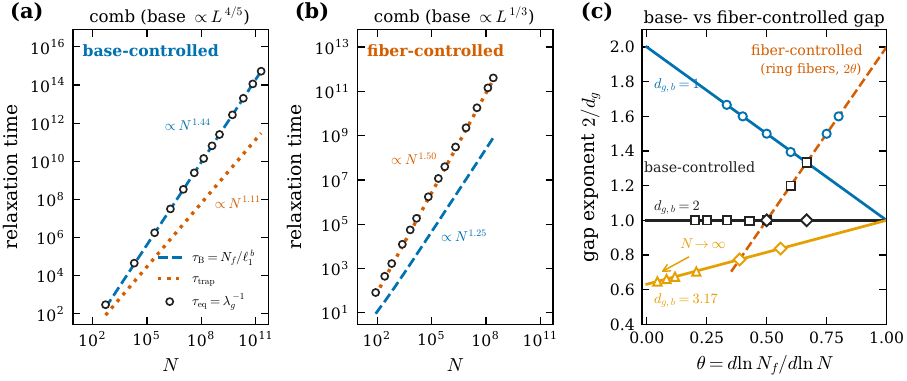}
\caption{Base- versus fiber-controlled equilibration. (a),(b) The two relaxation times of Eq.~\eqref{eq:series} for Dirac combs with ring fibers of length $L$ and a ring base of length $L^{4/5}$ (a) and $L^{1/3}$ (b): the diluted base time $\tauB=N_f/\ell^b_1$ (dashed) and the trapping time $\tauT$ (dotted). The exact equilibration time (circles) follows the time that grows faster with $N$: the base in (a), the fibers in (b). (c) Gap exponent $2/d_g$ as a function of the fiber share $\theta$ of the logarithmic mass growth. Base-controlled bundles lie on the straight lines $2(1-\theta)/d_{g,b}+\theta$ of Eq.~\eqref{eq:law}, shown for a ring base ($d_{g,b}=1$), bases at the fixed point $d_{g,b}=2$ (square lattice, Cayley tree), and the $(1,2)$-flower ($d_{g,b}=3.17$). For ring fibers the fiber-controlled exponent is $2\theta$ (dashed), and the gap follows the upper of the two lines, Eq.~\eqref{eq:criterion}. Symbols: local exponents at the largest size for combs (circles), brushes (squares), and small-world bases with Cayley fibers (diamonds); triangles show the $(1,2)$-flower with cubic-lattice fibers at generations $n=10$, $20$, $30$, and $59$, which move towards $\theta=0$ as $N$ grows (arrow).}
\label{fig:composition}
\end{figure*}

We now let the base and the fibers grow with a common length $L$. In the base-controlled regime, $\lambda_g \simeq \ell^b_1/N_f$ (Sec.~\ref{sec:Diffusion scales in composite networks}) and using $N=N_bN_f$ we find:
\begin{equation}
    \lambda_g\simeq\frac{\ell^b_1}{N_f}\sim N_b^{-\frac{2}{d_{g,b}}}N_f^{-1}
    =N^{-\frac{2}{d_{g,b}}}\,N_f^{\frac{2}{d_{g,b}}-1}.
    \label{eq:identity}
\end{equation}
This relation holds for any growth law. Only the share of $\ln N$ carried by the fibers matters, and it is measured by
\begin{equation}
    \theta\equiv\frac{\mathrm{d}\ln N_f}{\mathrm{d}\ln N}.
    \label{eq:theta}
\end{equation}
Differentiating Eq.~\eqref{eq:identity}, the effective Fiedler dimension, $d_g^{\rm eff}=-2\,\mathrm{d}\ln N/\mathrm{d}\ln\lambda_g$, obeys:
\begin{equation}
    \frac{1}{d_g^{\rm eff}}=\frac{1-\theta}{d_{g,b}}+\frac{\theta}{2}.
    \label{eq:law}
\end{equation}
The inverse Fiedler dimension interpolates linearly between the base value ($\theta=0$) and the value $1/2$ ($\theta=1$) imposed by the dilution factor $1/N_f$. For finite-dimensional components, $\theta=d_{f,f}/(d_{f,b}+d_{f,f})$, and Eq.~\eqref{eq:law} reduces to Eq.~\eqref{eq:old}. In terms of the \emph{gap exponent} $2/d_g$, Eq.~\eqref{eq:law} is a straight line in $\theta$ for every base [Fig.~\ref{fig:composition}(c)]: all lines end at $2/d_g=1$ for $\theta=1$, and the line of a base with $d_{g,b}=2$ is flat.
A base with $d_{g,b}=2$ is a fixed point of composition: $\lambda_g\sim N^{-1}$ for any fiber and any growth law, as long as the gap is base-controlled.

Let $r_\infty=\lim_{L\to\infty}(\mathrm{d}\ln N_b/\mathrm{d}\ln N_f)$ be the asymptotic ratio of the growth rates. Comparing $\tauB$ with Eq.~\eqref{eq:trap} gives the condition for a fiber-controlled gap,
\begin{equation}
    \frac{2}{d_{g,f}}-1>\frac{2\,r_\infty}{d_{g,b}} .
    \label{eq:criterion}
\end{equation}
For finite-dimensional components, $r_\infty=d_{f,b}/d_{f,f}$, and Eq.~\eqref{eq:criterion} is the condition derived in Ref.~\cite{Grimaldi_2026}. It also explains the Dirac comb whose base grows as $L^\alpha$: this comb is fiber-controlled for $\alpha<1/2$ and base-controlled otherwise~\cite{Grimaldi_2026}. In both cases the equilibration time follows whichever of the two times in Eq.~\eqref{eq:series} grows faster with $N$ [Fig.s~\ref{fig:composition}(a) and \ref{fig:composition}(b)].
 In the fiber-controlled phase, $d_g=d_{g,f}/\theta_\infty$.

\section{Four growth classes}
\label{sec:classes}

Eq.s~\eqref{eq:identity}--\eqref{eq:criterion} apply to any growth law. For each component we distinguish finite-dimensional growth, $N\sim L^{d_f}$, from small-world growth, $N\sim e^{\tilde d_fL}$, where $\tilde d_f$ is the transfinite fractal dimension of the transfractal~\cite{Rozenfeld_2007}. We use small-world in the scaling sense, $L\sim\ln N$, without implying the clustering of Watts-Strogatz networks~\cite{WattsStrogatz1998}. This yields four classes, summarized in Table~\ref{tab:classes} and tested in Fig.~\ref{fig:classes}. We compute the Fiedler eigenvalue exactly from Eq.~\eqref{eq:impurity}. This requires only $\ell^b_1$ and a single fiber, so we can reach bundles far larger than those accessible by direct diagonalization (Appendix~\ref{app:numerics}).

\begin{table}[h]
\begin{ruledtabular}
\begin{tabular}{lll}
Base / fiber & $\lambda_g$ & Fiber-controlled \\
\hline
f.d. / f.d. & $N^{-2/d_g}$, Eq.~\eqref{eq:old} & Eq.~\eqref{eq:criterion} \\
SW / f.d. & $N^{-2/d_{g,b}}(\ln N)^{-d_{f,f}(1-2/d_{g,b})}$ & never \\
f.d. / SW & $N^{-1}(\ln N)^{d_{f,b}(1-2/d_{g,b})}$ & never \\
SW / SW & $N^{-2/d_g}$, Eq.~\eqref{eq:ee} & never \\
\end{tabular}
\end{ruledtabular}
\caption{Fiedler eigenvalue of bundled networks in the four growth classes, for base-controlled equilibration. The last column gives the condition for a fiber-controlled gap. For the classes with a small-world component it assumes unweighted graphs, Eq.~\eqref{eq:dg2}, and a finite $d_{g,b}$ (expander bases are discussed in Sec.~\ref{trans+alg}). Finite-dimensional (f.d.) means $N\sim L^{d_f}$, small-world (SW) means $N\sim e^{\tilde d_fL}$.}
\label{tab:classes}
\end{table}

A general constraint applies to exponentially growing components. Mohar's inequality states that any connected graph with $N$ nodes and diameter $L$ satisfies $\lambda_g\ge4/(NL)$~\cite{Mohar1991}. For a small-world graph, $L\sim\ln N$, so $\lambda_g$ cannot decay faster than $1/(N\ln N)$. Therefore
\begin{equation}
    d_g\ge2\qquad\text{for any unweighted small-world graph}.
    \label{eq:dg2}
\end{equation}
A small-world component with $d_g<2$ would require couplings that weaken with scale.

\subsection{Finite-dimensional base and fibers}
\label{alg+alg}

Here $\theta=d_{f,f}/(d_{f,b}+d_{f,f})$ is constant, and Eq.~\eqref{eq:law} reproduces Eq.~\eqref{eq:old}. This is the only class with a fiber-controlled phase. Consider periodic combs and brushes~\cite{Grimaldi_2026} whose ring fibers have length $L$ and whose base (a ring or a square lattice) has side $L^\alpha$. The two regimes give
\begin{equation}
    d_g^{\rm comb}=\min\Big\{1+\alpha,\frac{2(1+\alpha)}{1+2\alpha}\Big\},\quad
    d_g^{\rm brush}=\min\{1+2\alpha,2\}.
    \label{eq:combs}
\end{equation}
Both architectures change regime at $\alpha=1/2$, where $d_g$ has a cusp [Fig.~\ref{fig:classes}(a)]. For $\alpha<1/2$ the base is too small to matter, and equilibration is set by the fibers, with $d_g=d_{g,f}/\theta$. For $\alpha>1/2$ the base controls the gap. The brush is then pinned at the fixed point $d_g=2$, while the comb follows Eq.~\eqref{eq:old}. The exact gaps, computed up to $N\simeq10^{19}$, reproduce Eq.~\eqref{eq:combs} on both sides of the cusp.

Deterministic fractals provide finite-dimensional bases with a nontrivial $d_{g,b}$. The $(u,v)$-flowers with $u\ge2$ are fractal, with $d_{f,b}=\ln(u+v)/\ln u$ and $d_{g,b}=2\ln(u+v)/\ln(uv)$~\cite{Rozenfeld_2007,Poggialini2025}. The $(2,2)$-flower is the diamond hierarchical lattice~\cite{BerkerOstlund1979}, a finite-dimensional realization of the fixed point $d_{g,b}=2$, whereas the $(2,3)$- and $(2,4)$-flowers have $d_{g,b}=1.80$ and $1.72$. With hypercubic fibers of side equal to the base diameter, $u^n$, the local Fiedler dimension of all nine bundles converges to Eq.~\eqref{eq:law}. Letting the ring fibers grow faster, with length $8^n$ on the $(2,3)$-flower of generation $n$, drives the bundle into the fiber-controlled phase predicted by Eq.~\eqref{eq:criterion}: $d_g$ converges to $d_{g,f}/\theta=1.774$ instead of the base-controlled value $1.906$ [Fig.~\ref{fig:classes}(f)].

\subsection{Small-world base, finite-dimensional fibers}
\label{trans+alg}

For $N_b\sim e^{\tilde d_{f,b}L}$ and $N_f\sim L^{d_{f,f}}$, $\theta\simeq d_{f,f}/(\tilde d_{f,b}L)\to0$, and Eq.~\eqref{eq:law} gives $d_g=d_{g,b}$. The naive limit $d_{f,b}\to\infty$ of Eq.~\eqref{eq:old} therefore yields the correct exponent, but it loses all information about the fibers. Since $L\simeq\ln N/\tilde d_{f,b}$, Eq.~\eqref{eq:identity} gives
\begin{equation}
    \lambda_g\sim N^{-\frac{2}{d_{g,b}}}\,(\ln N)^{-d_{f,f}\left(1-\frac{2}{d_{g,b}}\right)}.
    \label{eq:log}
\end{equation}
The fibers thus survive as a multiplicative logarithmic correction. Its exponent depends on $d_{f,f}$ but not on $d_{g,f}$, and by Eq.~\eqref{eq:dg2} it is never positive: finite-dimensional modules attached to a small-world core always slow down equilibration. For any finite $d_{g,b}$ the fiber-controlled phase disappears, because $\tauT$ grows only as a power of $\ln N$ ($r_\infty=\infty$). Convergence is slow. Inverting $N=L^{d_{f,f}}e^{\tilde d_{f,b}L}$ with the Lambert function~\cite{Corless1996} gives
\begin{equation}
    \frac{2}{d_g^{\rm eff}}=\frac{2}{d_{g,b}}+\frac{1-\frac{2}{d_{g,b}}}{1+W\!\big(\tfrac{\tilde d_{f,b}}{d_{f,f}}N^{\frac{1}{d_{f,f}}}\big)},
    \label{eq:lambert}
\end{equation}
with $W(x)\simeq\ln x$. For $d_{g,b}>2$, $d_g^{\rm eff}$ therefore approaches $d_{g,b}$ from below, with corrections of order $1/\ln N$. See Appendix~\ref{app:lambert} for the derivation.

We test this class with the $(1,2)$- and $(1,3)$-flowers~\cite{DGM2002,Rozenfeld_2007}, the $(1,2)$-tree, and the Cayley tree as bases. As fibers we use periodic hypercubic lattices with $d_{f,f}=1,2,3$ and side equal to the base diameter. For the flowers, the base gap follows from an exact decimation: a flower of generation $n+1$ is made of $1+v$ replicas of generation $n$ joined at their hubs, so the interior of each replica can be integrated out onto its two hubs, generation by generation, as for fractal and hierarchical lattices~\cite{Domany1983,BerkerOstlund1979,Rozenfeld_2007} (Appendix~\ref{app:numerics}). We thus reach generation $n=60$ and find $\ell^b_1\simeq12\cdot2^{-n}$ for the pseudofractal web and $\ell^b_1\simeq7.52\cdot3^{-n}$ for the $(1,3)$-flower. Hence $d_{g,b}=2\ln(1+v)/\ln v$, that is, $3.17$ and $2.52$, the values found for the Fiedler dimension of $(u,v)$-flowers in Ref.~\cite{Poggialini2025}: the base relaxation time scales as the first-passage time between hubs~\cite{Rozenfeld_2007}. For the flower bundles, the local dimension follows Eq.~\eqref{eq:law} over more than twenty-five decades in $N$ and stays below $d_{g,b}$ [Fig.~\ref{fig:classes}(b)]. A power-law fit at $N\simeq10^7$ gives $2.87$ instead of $3.17$, and even at $N\simeq10^{30}$ the local dimension is still $3.08$ for $d_{f,f}=3$. Trees instead have $N_b\ell^b_1\to{\rm const}$, i.e., they are small-world realizations of the fixed point $d_{g,b}=2$. For the Cayley tree the slowest mode is antisymmetric between two branches of the root and depends only on the distance from it inside each branch, so $\ell^b_1$ follows from a chain of $n$ shells at any depth $n$: we find $N_b\ell^b_1\to3/2$ up to depth $60$, consistent with $1/\lambda_g\propto N$ for Bethe lattices~\cite{Poggialini2025}. The $(1,2)$-tree is decimated like the flowers, since generation $n+1$ is made of one replica between the two hubs and one replica hanging from each of them; up to $n=60$, we find $N_b\ell^b_1\to2.751$. For them, $N\lambda_g$ converges to the same constant for every fiber, without any logarithm [Fig.~\ref{fig:classes}(c)].

The same classification applies to random small-world bases, whose Fiedler dimensions were measured in Ref.~\cite{Poggialini2025}. Barab\'asi-Albert trees ($m=1$) sit at the fixed point $d_{g,b}=2$, like the Cayley tree. Holme-Kim networks with triad-formation probability one~\cite{HolmeKim2002} are clustered and scale-free, yet they keep a finite $d_{g,b}$, between about $2.5$ and $3.6$, increasing with $m$~\cite{Poggialini2025}. Dressed with finite-dimensional modules they belong to this class, and Eq.~\eqref{eq:log} predicts a logarithmic slowdown of equilibration. Barab\'asi-Albert graphs with $m\ge2$ fall outside this classification: they are neither scale invariant nor transfractal, their Fiedler eigenvalue does not scale with size, and all their dimensions are infinite~\cite{Poggialini2025}. With such an expander base, $\tauB\sim N_f$, and recurrent fibers, $d_{g,f}<2$, control the gap through $\tauT\sim N_f^{2/d_{g,f}}$ irrespective of the growth rates. In both cases $\lambda_g$ decays only logarithmically in $N$ when the fiber size follows the base diameter.

\subsection{Finite-dimensional base, small-world fibers}
\label{alg+trans}

Now $N_f\sim e^{\tilde d_{f,f}L}$ dominates the mass, and $\theta\to1$. Eq.~\eqref{eq:identity} thus gives
\begin{equation}
    \lambda_g\sim N^{-1}\,(\ln N)^{d_{f,b}\left(1-\frac{2}{d_{g,b}}\right)}.
    \label{eq:AE}
\end{equation}
Small-world fibers therefore impose the universal value $d_g=2$, and the base survives only through the logarithm. A fiber-controlled gap would require $d_{g,f}<2$ (Eq.~\eqref{eq:criterion} with $r_\infty=0$), which Eq.~\eqref{eq:dg2} excludes. We attach Cayley trees of depth $L$ to ring, square, and cubic lattices of side $L$. These bases have $d_{f,b}=d_{g,b}=d$, so the logarithmic exponent is $d-2$. Fig.~\ref{fig:classes}(d) shows the predicted exponents $-1$, $0$, and $+1$. Depending on the base, the same fibers can thus make equilibration slower ($d=1$) or faster ($d=3$) than the pure $N^{-1}$ law.

\subsection{Small-world base and fibers}
\label{trans+trans}

\begin{figure*}[t]
\centering
\includegraphics[width=\textwidth]{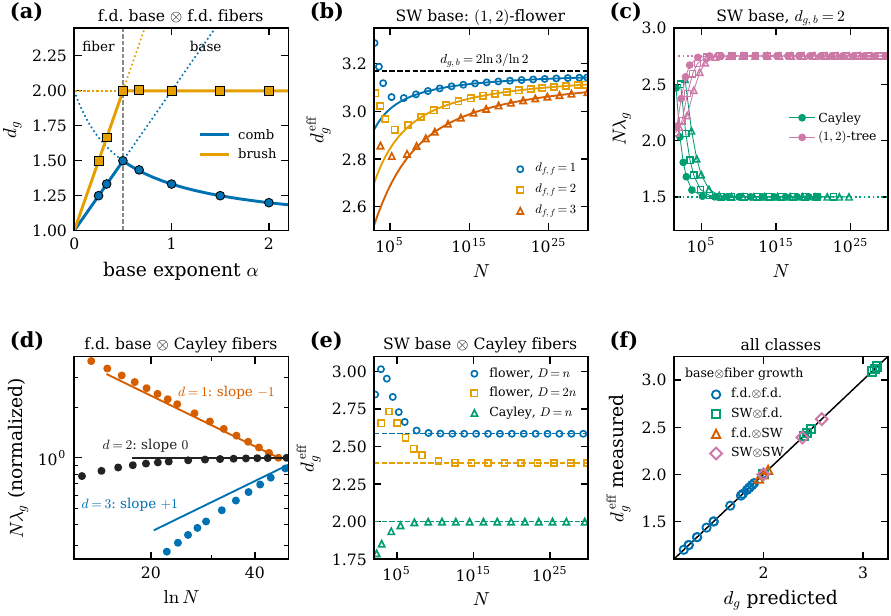}
\caption{Fiedler dimension in the four growth classes, from exact spectra. Panels (b), (c), and (e) show sizes up to $N=10^{30}$. (a) Finite-dimensional base and fibers: combs (ring base) and brushes (square-lattice base) of side $L^\alpha$ with ring fibers of length $L$. Lines: Eq.~\eqref{eq:combs}, with the non-selected regime dotted. Symbols: local $d_g$ at the largest size. (b) Small-world base with finite-dimensional fibers: $(1,2)$-flower with hypercubic fibers of dimension $d_{f,f}$. Symbols: local slopes; lines: Eq.~\eqref{eq:law}; dashed: $d_{g,b}$, approached from below. (c) Small-world bases at the fixed point $d_{g,b}=2$ (Cayley tree and $(1,2)$-tree): $N\lambda_g$ for Cayley-tree and $(1,2)$-tree bases with $d_{f,f}=1$ (filled), $2$, and $3$ (open) converges to $N_b\ell^b_1$ (dotted) for every fiber. (d) Finite-dimensional base and small-world fibers: $d$-dimensional lattices of side $L$ with Cayley fibers of depth $L$. $N\lambda_g$ is normalized at the largest size, and lines show the slopes $d-2$ predicted by Eq.~\eqref{eq:AE}. (e) Small-world base and fibers: $(1,2)$-flower with Cayley fibers of depth $D=n$ and $2n$, and Cayley base with Cayley fibers. Dashed lines: Eq.~\eqref{eq:ee}. (f) Measured versus predicted $d_g$ for all $44$ architectures, including fractal $(u,v)$-flower bases with $u\ge2$.}
\label{fig:classes}
\end{figure*}

When both sectors are small-world, $\theta_\infty=\tilde d_{f,f}/(\tilde d_{f,b}+\tilde d_{f,f})$ is finite, and Eq.~\eqref{eq:law} gives
\begin{equation}
    d_g=2\,\frac{\tilde d_{f,f}+\tilde d_{f,b}}{\tilde d_{f,f}+2\tilde d_{f,b}/d_{g,b}} .
    \label{eq:ee}
\end{equation}
This is Eq.~\eqref{eq:old} with transfinite dimensions in place of fractal ones. Unlike the mixed classes, the two sectors now combine into a genuinely new exponent. As in the previous class, a fiber-controlled gap is excluded for unweighted fibers. For $(1,2)$-flower bases of generation $n$ with Cayley fibers of depth $D=n$ and $D=2n$, Eq.~\eqref{eq:ee} predicts $d_g=2.585$ and $2.390$, the values reached by the exact spectra [Fig.~\ref{fig:classes}(e)]. With a Cayley base, the fixed point gives $d_g=2$ for any fiber depth.

Fig.~\ref{fig:classes}(f) collects the local Fiedler dimension at the largest size for all $44$ architectures studied, spanning the four classes and $1.2\le d_g\le3.1$. In every case, it falls on the prediction of Eq.s~\eqref{eq:law} and \eqref{eq:criterion}, evaluated with the actual growth laws at the same size. We infer the following general rule: the sector that dominates the growth of $\ln N$ fixes the exponent, the other survives only as a logarithmic correction, and the exponents mix only when both sectors grow at the same rate. The same logic covers ultrasmall bases, with $L\sim\ln\ln N$~\cite{Cohen2003}, where the correction in Eq.~\eqref{eq:log} becomes $(\ln\ln N)^{-d_{f,f}(1-2/d_{g,b})}$. For bases at the fixed point, such as Barab\'asi-Albert trees with $d_{g,b}=2$~\cite{Poggialini2025,Grimaldi_2026}, the correction vanishes altogether.

\section{Outlook}

We asked which level of a modular network, the modules or their wiring, sets the slowest time scale. For bundled networks, the answer rests on a series law for relaxation times, which also delimits the validity of the perturbative spectrum used in Ref.~\cite{Grimaldi_2026}, and it is controlled by a single quantity, the share of the logarithmic mass growth carried by each level. In scale-invariant architectures, the modules can control equilibration only when both levels are finite-dimensional and the modules are large enough (Eq.~\eqref{eq:criterion}). When a small-world core carries finite-dimensional modules, it imposes its Fiedler dimension, which is never smaller than two, and the modules leave only a logarithmic trace. When the modules are the small-world level, $d_g=2$ and the core leaves the trace. When both are small-world, transfinite dimensions combine as fractal dimensions do.

These results have three consequences for network science. First, effective dimensions extracted from finite data on modular small-world systems, such as brain networks~\cite{Peach2022,Poggialini2025}, can be biased by logarithmic corrections that vanish only at astronomically large sizes, and should be read as effective exponents. Second, finite-dimensional modules attached to a small-world core always slow down equilibration, and hence consensus and synchronization~\cite{Pecora1998,Barahona2002}, by a factor that grows as a power of $\ln N$; only cores at the fixed point $d_g=2$, such as Cayley and Barab\'asi-Albert trees, are insensitive to the modules beyond their mass. Third, the competition between the two times of the series law is reminiscent of the abrupt change of $\lambda_g$ observed when two networks are coupled~\cite{RadicchiArenas2013}, where the inter-layer coupling plays the role of the base. Natural extensions include hierarchical architectures, where the same argument applied level by level would yield a hierarchy of relaxation times, interconnected and multiplex networks~\cite{Gomez2013}, bundles with random fibers~\cite{Jonsson_2008,Destri2002}, and weighted small-world cores, the only route to $d_{g,b}<2$ in the small-world class.

\begin{acknowledgments}
We are deeply grateful to R.\,Burioni, D.\,Cassi, and A.\,Vezzani for many insightful discussions and for their collaboration on the previous work that motivated this study. We also thank O.\,Falconi and L.\,Lucarini for useful discussions. LG would like to thank the École Normale Supérieure for its hospitality during the initial stages of this work. PV acknowledges the Spanish Ministry of Research and Innovation and Agencia Estatal de Investigaci\'on (AEI), MICIN/AEI/10.13039/501100011033, for financial support through Project No.~PID2023-149174NB-I00, also funded by European Regional Development Funds, and Ref.~PID2020-113681GB-I00.

\end{acknowledgments}

\appendix
\section{The spectral gap of a bundle}
\label{app:series}

In the basis that diagonalizes the base, Eq.~\eqref{eq:impurity} is a single fiber carrying a repulsive impurity of strength $\ell^b_{n_1}$ on its root. The impurity pushes the levels upwards, since $\partial\lambda/\partial\ell^b_{n_1}=|\tilde\psi(0)|^2\ge0$, but it acts on one site out of $N_f$, and it cannot push the lowest level of a sector above the first excited level $\lambda_{1,f}$ of the free fiber, which the sector $n_1=0$ already contributes to the spectrum. The gap of the bundle is therefore the lowest level of the sector $n_1=1$, for any base and any fiber, and the base- and the fiber-controlled regimes are the weak- and the strong-coupling limits of the same state.

This is the standard one-impurity problem. Writing Eq.~\eqref{eq:impurity} as $(\hat L^f-\lambda)\tilde\psi=-\ell^b_1\tilde\psi(0)|0\rangle$ and evaluating the solution at the root, a level outside the fiber spectrum must satisfy $1+\ell^b_1G(\lambda)=0$, where $G(\lambda)=\langle0|(\hat L^f-\lambda)^{-1}|0\rangle=\sum_kw_k/(\lambda_{k,f}-\lambda)$ is the fiber propagator at the root and $w_k=|\phi_k(0)|^2$ the weight there of the fiber mode $\phi_k$. The uniform mode has $\lambda_{0,f}=0$ and $w_0=1/N_f$; separating it from the rest and multiplying by $N_f/\ell^b_1$,
\begin{equation}
    \frac{1}{\lambda}=\frac{N_f}{\ell^b_1}+N_fS(\lambda),
    \qquad
    S(\lambda)=\sum_{k\ge1}\frac{w_k}{\lambda_{k,f}-\lambda}.
    \label{eq:secular}
\end{equation}
The first term is the relaxation time of the base, diluted by the fibers; the second is the response of the internal modes of the fiber.

Since the level we are after lies below $\lambda_{1,f}$, we can bound $S$ between its static value $R_0\equiv S(0)=(\hat L^{f+})_{00}$ and $R_0/(1-\lambda/\lambda_{1,f})$. At $\lambda=\lambda_g$ this gives
\begin{equation}
    \tauB+\tauT\;\le\;\frac{1}{\lambda_g}\;\le\;\tauB+\frac{\tauT}{1-\frac{\lambda_g}{\lambda_{1,f}}},
    \qquad \tauT=N_fR_0,
    \label{eq:bounds}
\end{equation}
that is, Eq.~\eqref{eq:series}, which becomes exact as soon as the bundle relaxes more slowly than an isolated fiber. The second time is a trapping time: the mean first-passage time to the root solves $(\hat L^fh)_i=1$ for $i\ne0$ with $h_0=0$, so that $h=-N_f\hat L^{f+}|0\rangle+{\rm const}$, whose average over a uniform starting point is $N_fR_0$. For a ring $\tauT=(N_f^2-1)/12$, and for a chain grafted at one end $\tauT=(N_f-1)(2N_f-1)/6$.

In the opposite limit, $\ell^b_1R_0\gg1$, the root is frozen and $\lambda_g$ saturates at the gap $\lambda_D$ of a fiber with a pinned root, so that the plateau of Fig.~\ref{fig:bundles}(c) is $\tauT\lambda_D$. In one dimension both factors are elementary. Pinning one site of a ring leaves a chain of $N_f-1$ sites with both ends fixed, $\lambda_D=2[1-\cos(\pi/N_f)]\simeq\pi^2/N_f^2$, which together with $\tauT\simeq N_f^2/12$ gives $\tauT\lambda_D\to\pi^2/12\simeq0.822$. Pinning the grafted end of a chain leaves one end fixed and one free, $\lambda_D=2\{1-\cos[\pi/(2N_f-1)]\}\simeq\pi^2/(4N_f^2)$, four times smaller, while $\tauT\simeq N_f^2/3$ is four times larger, so that the plateau is again $\pi^2/12$. It is thus a property of the fiber alone, and it tends to unity for transient fibers, where pinning the root barely deforms the uniform mode, $\lambda_D\ll\lambda_{1,f}$, and Eq.~\eqref{eq:secular} keeps its one-pole form.

The numerical procedures used to evaluate Eq.~\eqref{eq:secular} for the bundles of Fig.s~\ref{fig:bundles}--\ref{fig:classes} are described in Appendix~\ref{app:numerics}.

\section{Exact Lambert-$W$ solution for a small-world base with finite-dimensional fibers}
\label{app:lambert}

Consider
\begin{equation}
    N=L^a e^{bL},
    \qquad a>0,\;b>0.
\label{eq:app-N}
\end{equation}
In order to invert Eq.~\eqref{eq:app-N}, we use the principal Lambert function $W(z)$, defined through $W(z)e^{W(z)}=z$. Since the argument is positive, only the single real principal branch is involved. Therefore
\begin{equation}
    L=\frac{a}{b}
    W\!\left(\frac{b}{a}N^{\frac{1}{a}}\right).
\label{eq:app-L}
\end{equation}
In the following, we will consider the argument of the Lambert $W$ to always be $z = (b/a) N^{1/a}$.

For a base gap exponent $\alpha_b$, we have $\lambda_b\sim L^{-a}e^{-\alpha_b bL}$. Substituting Eq.~\eqref{eq:app-L}, and absorbing $N$-independent constants, we find
\begin{equation}
    \lambda_b \propto W^{-a}e^{-\alpha_b aW}.
\label{eq:app-pre}
\end{equation}
Using the exact identity $e^W=z/W$, we obtain $e^{-\alpha_b aW}=(W/z)^{\alpha_b a}$, and thus
\begin{align}
    \lambda_b \propto z^{-\alpha_b a}W^{a(\alpha_b-1)}.
\end{align}
Plugging back in the definition of $z$, we finally obtain
\begin{equation}
    \lambda_b
    \propto
    N^{-\alpha_b}
    W\!\left(\frac{b}{a}N^{\frac{1}{a}}\right)^{a(\alpha_b-1)}.
\label{eq:app-exact}
\end{equation}
For large positive $z$, the Lambert $W$ displays the asymptotic behavior
\begin{equation}
    W(z)=\ln z-\ln\ln z
    +O\!\left(\frac{\ln\ln z}{\ln z}\right).
\end{equation}
Hence, Eq.~\eqref{eq:app-exact} becomes
\begin{equation}
    \lambda_b
    \sim
    N^{-\alpha_b}
    (\ln N)^{a(\alpha_b-1)}
    \left[1+O\!\left(\frac{\ln\ln N}{\ln N}\right)\right],
\label{eq:app-asym}
\end{equation}
which is exactly Eq.~\eqref{eq:log}, upon suitable matching of the exponents with the dimensions of each sector.

The same reasoning can be applied to the composition of a finite-dimensional base and small-world fibers, leading to the same result as Subsection~\ref{alg+trans} upon suitable identification of the parameters in Eq.~\eqref{eq:app-N} with the scaling exponents of the individual components.

\section{Numerical methods}
\label{app:numerics}

None of the Fiedler eigenvalues shown in Fig.s~\ref{fig:bundles}--\ref{fig:classes} requires diagonalizing the bundle. By Eq.~\eqref{eq:impurity} and Appendix~\ref{app:series}, $\lambda_g$ is the lowest level of a single fiber carrying the impurity $\ell^b_1$ on its root. Two quantities are therefore needed: the gap $\ell^b_1$ of the base, and the lowest level of the fiber with the impurity. For bundles with up to $N\sim10^6$ nodes, we checked every step against sparse shift-invert diagonalization of the full Laplacian~\cite{ARPACK,SciPy}. Whenever $\lambda_g$ becomes very small, as for large small-world bases, we use arbitrary-precision arithmetic with $50$ to $90$ significant digits~\cite{mpmath}.

\emph{Gap of the base.} Rings and periodic square and cubic lattices of side $L$ have plane-wave modes and $\ell^b_1=2[1-\cos(2\pi/L)]$. For the Cayley tree, the slowest mode is antisymmetric between two branches of the root and depends only on the distance from it inside each branch~\cite{CaiChen1997,RojoSoto2005}. It is then the lowest level of a chain of $n$ shells with the root fixed, with couplings $\sqrt2$ between consecutive shells, which we locate by counting sign changes of the Sturm sequence and bisecting in $\lambda$~\cite{Barth1967}. The flowers and the $(1,2)$-tree are treated by exact decimation, the method introduced for the spectra of deterministic fractals~\cite{Domany1983,Rammal1984} and applied to the pseudofractal web in Ref.~\cite{Bollt2005}. Generation $n+1$ consists of replicas of generation $n$ that share only their two hubs. At a trial value of $\lambda$, the interior of each replica can therefore be eliminated from $\hat L-\lambda$ by Gaussian elimination, leaving a $2\times2$ effective coupling between its hubs. The couplings of generation $n+1$ follow from those of generation $n$ by adding replicas in parallel and by eliminating the intermediate hubs of replicas in series, or the free hub of a hanging replica in the $(1,2)$-tree, as in the renormalization of hierarchical lattices~\cite{BerkerOstlund1979}. By Sylvester's law of inertia~\cite{GolubVanLoan}, the number of negative pivots met along this elimination equals the number of eigenvalues of $\hat L$ below $\lambda$, as in Dean's negative-eigenvalue counting for disordered harmonic chains~\cite{Dean1972}. A bisection in $\lambda$ then gives $\ell^b_1$ with a number of operations proportional to $n$ rather than to $N_b$, up to $n=60$. The recursion extends to any $(u,v)$-flower, whose generation $n+1$ consists of a chain of $u$ and a chain of $v$ replicas in parallel. It reproduces the sparse-diagonalization values at all generations where both are available, including the Fiedler eigenvalues of the $(1,2)$-, $(1,3)$-, $(1,4)$-, $(2,2)$-, $(2,3)$-, and $(2,4)$-flowers obtained by direct diagonalization up to $N\simeq3\times10^6$ in Ref.~\cite{Poggialini2025}, to all the digits given there. For the $(1,3)$-flower in Fig.~\ref{fig:classes}(f) we use sparse diagonalization.

\emph{Fiber with an impurity.} For ring fibers of length $N_f$, Eq.~\eqref{eq:secular} reduces to $2\sin\omega=\ell^b_1\cot(N_f\omega/2)$, with $\lambda_g=4\sin^2(\omega/2)$~\cite{Grimaldi_2026}, whose lowest root is found by bisection. Periodic hypercubic fibers have plane-wave modes, all with weight $w_k=1/N_f$ on the root, so that $S(\lambda)$ in Eq.~\eqref{eq:secular} is a lattice Green function, i.e., a sum over the Brillouin zone~\cite{Montroll1965}. For the small-world bases, $\lambda_g\ll\lambda_{1,f}$, and we keep the first two terms of its expansion in $\lambda$, both summed exactly, the neglected ones being of relative order $(\lambda_g/\lambda_{1,f})^2$; for smaller bases we diagonalize the fiber directly. For Cayley fibers, the lowest level is nondegenerate and positive, hence invariant under the rotations that fix the root. It is thus the lowest level of a chain of $D+1$ shells carrying the impurity on the first one, located again by Sturm counting.

\emph{Local exponents.} The local Fiedler dimension in Fig.s~\ref{fig:composition} and \ref{fig:classes} is $d_g^{\rm eff}=-2\,\Delta\ln N/\Delta\ln\lambda_g$ between consecutive sizes of a family. The fiber share $\theta=\Delta\ln N_f/\Delta\ln N$ is evaluated on the same pair, and the predictions of Eq.s~\eqref{eq:law} and \eqref{eq:criterion} in Fig.~\ref{fig:classes}(f) use this value.

\bibliography{sample}

@article{Jacob1977,
  title={Evolution and tinkering},
  author={Jacob, Fran{\c{c}}ois},
  journal={Science},
  volume={196},
  number={4295},
  pages={1161--1166},
  year={1977},
  doi={10.1126/science.860134},
  publisher={American Association for the Advancement of Science}
}

@article{Boson_comb2,
doi = {10.1088/0953-4075/34/23/314},
url = {https://dx.doi.org/10.1088/0953-4075/34/23/314},
year = {2001},
month = {nov},
publisher = {},
volume = {34},
number = {23},
pages = {4697},
author = {R Burioni and D Cassi and M Rasetti and P Sodano and A Vezzani},
title = {{Bose-Einstein condensation on inhomogeneous complex
networks}},
journal = {J. Phys. B: At. Mol. Opt. Phys.}
}

@article{spectral_dim_bund,
author = {Cassi, Davide and Regina, Sofia},
title = {Diffusion and Harmonic Oscillations on Bundled Structures: Analytical Techniques and Dynamical Dimension Splitting},
journal = {Mod. Phys. Lett. B},
volume = {11},
number = {23},
pages = {997--1011},
year = {1997},
doi = {10.1142/S0217984997001213}
}

@article{spectral_fiedler_instability,
doi = {10.1209/epl/i2002-00445-5},
url = {https://dx.doi.org/10.1209/epl/i2002-00445-5},
year = {2002},
month = {jun},
publisher = {},
volume = {58},
number = {6},
pages = {806},
author = {R. Burioni and D. Cassi and M. P. Fontana and A. Vulpiani},
title = {Vibrational thermodynamic instability  of recursive networks},
journal = {Europhys. Lett.}
}

@article{RW_review,
doi = {10.1088/0305-4470/38/8/R01},
url = {https://dx.doi.org/10.1088/0305-4470/38/8/R01},
year = {2005},
month = {feb},
publisher = {},
volume = {38},
number = {8},
pages = {R45},
author = {Burioni, R and Cassi, D},
title = {Random walks on graphs: ideas, techniques and results},
journal = {J. Phys. A: Math. Gen.}
}

@article{Crisp2015,
  author       = {Crisp, Andrew and Boschetti, Chiara and Perry, Michael and others},
  title        = {Expression of multiple horizontally acquired genes is a hallmark of both vertebrate and invertebrate genomes},
  journal      = {Genome Biol.},
  year         = {2015},
  volume       = {16},
  pages        = {50},
  doi          = {10.1186/s13059-015-0607-3}
}

@article{Rammal1984,
  author       = {R. Rammal},
  title        = {Spectrum of harmonic excitations on fractals},
  journal      = {J. Phys. (Paris)},
  volume       = {45},
  number       = {2},
  pages        = {191--206},
  year         = {1984},
  doi          = {10.1051/jphys:01984004502019100}
}

@article{Pace2008,
author = {John K. Pace  and Cl{\'e}ment Gilbert  and Marlena S. Clark  and C{\'e}dric Feschotte },
title = {{Repeated horizontal transfer of a DNA transposon in mammals and other tetrapods}},
journal = {Proc. Natl. Acad. Sci. U.S.A.},
volume = {105},
number = {44},
pages = {17023--17028},
year = {2008},
doi = {10.1073/pnas.0806548105}
}

@article{Grillner2016,
  author       = {Grillner, Sten and Robertson, Brita},
  title        = {The Basal Ganglia Over 500 Million Years},
  journal      = {Curr. Biol.},
  year         = {2016},
  volume       = {26},
  number       = {20},
  pages        = {R1088--R1100},
  month        = {October},
  doi          = {10.1016/j.cub.2016.06.041}
}

@article{Villegas2020,
  author       = {Villegas, Pablo and Mu{\~n}oz, Miguel A. and Bonachela, Juan A.},
  title        = {{Evolution in the Debian GNU/Linux software network: analogies and differences with gene regulatory networks}},
  journal      = {J. R. Soc. Interface},
  year         = {2020},
  volume       = {17},
  number       = {163},
  pages        = {20190845},
  month        = {February},
  doi          = {10.1098/rsif.2019.0845}
}

@article{Fortuna2011,
author = {Miguel A. Fortuna  and Juan A. Bonachela  and Simon A. Levin },
title = {Evolution of a modular software network},
journal = {Proc. Natl. Acad. Sci. U.S.A.},
volume = {108},
number = {50},
pages = {19985--19989},
year = {2011},
doi = {10.1073/pnas.1115960108}}

@article{AlexanderOrbach,
author = {S. Alexander and R. Orbach },
title = {Density of states on fractals: ``fractons''},
journal = {J. Phys. (Paris) Lett.},
volume = {43},
number = {17},
pages = {L625--L631},
year = {1982},
doi = {10.1051/jphyslet:019820043017062500}
}

@article{Fiedler_eig,
author = {M. Fiedler},
title = {Algebraic connectivity of graphs},
journal = {Czech. Math. J.},
volume = {23},
number = {2},
pages = {298--305},
year = {1973},
doi = {10.21136/CMJ.1973.101168}
}

@article{comb_of_comb,
    author = {Zhu, Yuan and Yuan, Zhenhua and Peng, Junhao},
    title = {First passage properties of d-dimensional finite combs with different growth modes},
    journal = {Chaos},
    volume = {35},
    number = {6},
    pages = {063120},
    year = {2025},
    month = {06},
    issn = {1054-1500},
    doi = {10.1063/5.0273216},
    url = {https://doi.org/10.1063/5.0273216},
}

@article{Poggialini2025,
  title = {Networks with Many Structural Scales: A Renormalization Group Perspective},
  author = {Poggialini, Anna and Villegas, Pablo and Mu\~noz, Miguel A. and Gabrielli, Andrea},
  journal = {Phys. Rev. Lett.},
  volume = {134},
  issue = {5},
  pages = {057401},
  numpages = {6},
  year = {2025},
  month = {Feb},
  publisher = {American Physical Society},
  doi = {10.1103/PhysRevLett.134.057401}
}

@article{Peach2022,
  author       = {Peach, Robert and Arnaudon, Alexis and Barahona, Mauricio},
  title        = {Relative, local and global dimension in complex networks},
  journal      = {Nat. Commun.},
  year         = {2022},
  volume       = {13},
  pages        = {3088},
  month        = {June},
  doi          = {10.1038/s41467-022-30705-w}
}

@article{Villegas2025,
  title = {Strange attractors in complex networks},
  author = {Villegas, Pablo},
  journal = {Phys. Rev. E},
  volume = {111},
  issue = {4},
  pages = {L042301},
  numpages = {5},
  year = {2025},
  month = {Apr},
  publisher = {American Physical Society},
  doi = {10.1103/PhysRevE.111.L042301},
  url = {https://link.aps.org/doi/10.1103/PhysRevE.111.L042301}
}

@article{Cassi1992,
  title = {{Phase transitions and random walks on graphs: A generalization of the Mermin-Wagner theorem to disordered lattices, fractals, and other discrete structures}},
  author = {Cassi, Davide},
  journal = {Phys. Rev. Lett.},
  volume = {68},
  issue = {24},
  pages = {3631--3634},
  numpages = {0},
  year = {1992},
  month = {Jun},
  publisher = {American Physical Society},
  doi = {10.1103/PhysRevLett.68.3631}
}

@article{Burioni1996,
  title = {Universal Properties of Spectral Dimension},
  author = {Burioni, Raffaella and Cassi, Davide},
  journal = {Phys. Rev. Lett.},
  volume = {76},
  issue = {7},
  pages = {1091--1093},
  numpages = {0},
  year = {1996},
  month = {Feb},
  publisher = {American Physical Society},
  doi = {10.1103/PhysRevLett.76.1091}
}

@article{Cassi1996_Bundled,
  title = {Random Walks on Bundled Structures},
  author = {Cassi, Davide and Regina, Sofia},
  journal = {Phys. Rev. Lett.},
  volume = {76},
  issue = {16},
  pages = {2914--2917},
  numpages = {0},
  year = {1996},
  month = {Apr},
  publisher = {American Physical Society},
  doi = {10.1103/PhysRevLett.76.2914}
}

@article{Jonsson_2008,
   title={The spectral dimension of random brushes},
   volume={41},
   ISSN={1751-8121},
   url={http://dx.doi.org/10.1088/1751-8113/41/4/045005},
   DOI={10.1088/1751-8113/41/4/045005},
   number={4},
   journal={J. Phys. A: Math. Theor.},
   publisher={IOP Publishing},
   author={Jonsson, Thordur and Stef{\'a}nsson, Sigurdur Orn},
   year={2008},
   month=jan, pages={045005} }

@article{Hattori1987,
  author       = {Hattori, Kumiko and Hattori, Tetsuya and Watanabe, Hiroshi},
  title        = {Gaussian Field Theories on General Networks and the Spectral Dimensions},
  journal      = {Prog. Theor. Phys. Suppl.},
  volume       = {92},
  pages        = {108--143},
  year         = {1987},
  issn         = {0375-9687},
  doi          = {10.1143/PTPS.92.108}
}

@article{LRG,
  title={{Laplacian Renormalization Group for heterogeneous networks}},
  author={Villegas, Pablo and Gili, Tommaso and Caldarelli, Guido and Gabrielli, Andrea},
  journal={Nat. Phys.},
  volume={19},
  number={3},
  pages={445--450},
  year={2023},
  doi={10.1038/s41567-022-01866-8}
}

@article{Modularity,
  title = {{Multi-scale Laplacian community detection in heterogeneous networks}},
  author = {Villegas, Pablo and Gabrielli, Andrea and Poggialini, Anna and Gili, Tommaso},
  journal = {Phys. Rev. Res.},
  volume = {7},
  issue = {1},
  pages = {013065},
  numpages = {16},
  year = {2025},
  month = {Jan},
  publisher = {American Physical Society},
  doi = {10.1103/PhysRevResearch.7.013065}
}

@article{InfoCore,
  title = {Laplacian paths in complex networks: Information core emerges from entropic transitions},
  author = {Villegas, Pablo and Gabrielli, Andrea and Santucci, Francesca and Caldarelli, Guido and Gili, Tommaso},
  journal = {Phys. Rev. Res.},
  volume = {4},
  issue = {3},
  pages = {033196},
  numpages = {12},
  year = {2022},
  month = {Sep},
  publisher = {American Physical Society},
  doi = {10.1103/PhysRevResearch.4.033196}
}

@article{Domenico2016,
  title = {Spectral Entropies as Information-Theoretic Tools for Complex Network Comparison},
  author = {De Domenico, Manlio and Biamonte, Jacob},
  journal = {Phys. Rev. X},
  volume = {6},
  issue = {4},
  pages = {041062},
  numpages = {16},
  year = {2016},
  month = {Dec},
  publisher = {American Physical Society},
  doi = {10.1103/PhysRevX.6.041062}
}

@article{Destri2002,
doi = {10.1088/0305-4470/35/45/301},
year = {2002},
month = {oct},
publisher = {},
volume = {35},
number = {45},
pages = {9499},
author = {Claudio Destri and Luca Donetti},
title = {The spectral dimension of random trees},
journal = {J. Phys. A: Math. Gen.}

}

@article{Pecora1998,
  title = {Master Stability Functions for Synchronized Coupled Systems},
  author = {Pecora, Louis M. and Carroll, Thomas L.},
  journal = {Phys. Rev. Lett.},
  volume = {80},
  issue = {10},
  pages = {2109--2112},
  numpages = {0},
  year = {1998},
  month = {Mar},
  publisher = {American Physical Society},
  doi = {10.1103/PhysRevLett.80.2109},
  url = {https://link.aps.org/doi/10.1103/PhysRevLett.80.2109}
}

@article{Barahona2002,
  title = {Synchronization in Small-World Systems},
  author = {Barahona, Mauricio and Pecora, Louis M.},
  journal = {Phys. Rev. Lett.},
  volume = {89},
  issue = {5},
  pages = {054101},
  numpages = {4},
  year = {2002},
  month = {Jul},
  publisher = {American Physical Society},
  doi = {10.1103/PhysRevLett.89.054101},
  url = {https://link.aps.org/doi/10.1103/PhysRevLett.89.054101}
}

@article{Grimaldi_2026,
   title={Asymptotic versus mesoscopic spectral dimensions in networks and inhomogeneous structures},
   volume={114},
   ISSN={2470-0053},
   url={http://dx.doi.org/10.1103/zdpg-dr6t},
   DOI={10.1103/zdpg-dr6t},
   number={2},
   journal={Phys. Rev. E},
   publisher={American Physical Society (APS)},
   author={Grimaldi, Lorenzo and Villegas, Pablo and Vezzani, Alessandro and Burioni, Raffaella and Cassi, Davide and Gabrielli, Andrea},
   year={2026},
   month=Aug, pages={L022301} }

@article{Rozenfeld_2007,
   title={Fractal and transfractal recursive scale-free nets},
   volume={9},
   ISSN={1367-2630},
   url={http://dx.doi.org/10.1088/1367-2630/9/6/175},
   DOI={10.1088/1367-2630/9/6/175},
   number={6},
   journal={New J. Phys.},
   publisher={IOP Publishing},
   author={Rozenfeld, Hern{\'a}n D and Havlin, Shlomo and ben-Avraham, Daniel},
   year={2007},
   month=jun, pages={175} }

@article{Corless1996,
author = {Corless, Robert and Gonnet, Gaston and Hare, D. and Jeffrey, David and Knuth, D.},
year = {1996},
month = {01},
pages = {329--359},
title = {{On the Lambert W Function}},
volume = {5},
journal = {Adv. Comput. Math.},
doi = {10.1007/BF02124750}
}

@article{Mohar1991,
  author  = {Mohar, Bojan},
  title   = {Eigenvalues, diameter, and mean distance in graphs},
  journal = {Graphs Comb.},
  volume  = {7},
  pages   = {53--64},
  year    = {1991},
  doi     = {10.1007/BF01789463}
}

@article{DGM2002,
  title   = {Pseudofractal scale-free web},
  author  = {Dorogovtsev, S. N. and Goltsev, A. V. and Mendes, J. F. F.},
  journal = {Phys. Rev. E},
  volume  = {65},
  pages   = {066122},
  year    = {2002},
  doi     = {10.1103/PhysRevE.65.066122}
}

@article{WattsStrogatz1998,
  title   = {Collective dynamics of `small-world' networks},
  author  = {Watts, Duncan J. and Strogatz, Steven H.},
  journal = {Nature},
  volume  = {393},
  pages   = {440--442},
  year    = {1998},
  doi     = {10.1038/30918}
}

@article{Cohen2003,
  title   = {Scale-Free Networks Are Ultrasmall},
  author  = {Cohen, Reuven and Havlin, Shlomo},
  journal = {Phys. Rev. Lett.},
  volume  = {90},
  pages   = {058701},
  year    = {2003},
  doi     = {10.1103/PhysRevLett.90.058701}
}

@article{DorogovtsevRMP2008,
  title   = {Critical phenomena in complex networks},
  author  = {Dorogovtsev, S. N. and Goltsev, A. V. and Mendes, J. F. F.},
  journal = {Rev. Mod. Phys.},
  volume  = {80},
  pages   = {1275--1335},
  year    = {2008},
  doi     = {10.1103/RevModPhys.80.1275}
}

@article{Condamin2007,
  title   = {First-passage times in complex scale-invariant media},
  author  = {Condamin, S. and B{\'e}nichou, O. and Tejedor, V. and Voituriez, R. and Klafter, J.},
  journal = {Nature},
  volume  = {450},
  pages   = {77--80},
  year    = {2007},
  doi     = {10.1038/nature06201}
}

@article{CaiChen1997,
  title   = {Rouse Dynamics of a Dendrimer Model in the $\vartheta$ Condition},
  author  = {Cai, Chunhua and Chen, Zheng Yu},
  journal = {Macromolecules},
  volume  = {30},
  pages   = {5104--5117},
  year    = {1997},
  doi     = {10.1021/ma970059z}
}

@article{HolmeKim2002,
  title = {Growing scale-free networks with tunable clustering},
  author = {Holme, Petter and Kim, Beom Jun},
  journal = {Phys. Rev. E},
  volume = {65},
  pages = {026107},
  year = {2002},
  doi = {10.1103/PhysRevE.65.026107}
}

@book{ARPACK,
  author    = {Lehoucq, R. B. and Sorensen, D. C. and Yang, C.},
  title     = {{ARPACK} Users' Guide: Solution of Large-Scale Eigenvalue Problems with Implicitly Restarted {Arnoldi} Methods},
  publisher = {SIAM},
  address   = {Philadelphia},
  year      = {1998},
  doi       = {10.1137/1.9780898719628}
}

@article{SciPy,
  author  = {Virtanen, Pauli and others},
  title   = {{SciPy} 1.0: fundamental algorithms for scientific computing in {Python}},
  journal = {Nat. Methods},
  volume  = {17},
  pages   = {261--272},
  year    = {2020},
  doi     = {10.1038/s41592-019-0686-2}
}

@misc{mpmath,
  author = {{The mpmath development team}},
  title  = {mpmath: a {Python} library for arbitrary-precision floating-point arithmetic (version 1.3.0)},
  note   = {\url{https://mpmath.org/}},
  year   = {2023}
}

@article{RojoSoto2005,
  author  = {Rojo, Oscar and Soto, Ricardo},
  title   = {The spectra of the adjacency matrix and {Laplacian} matrix for some balanced trees},
  journal = {Linear Algebra Appl.},
  volume  = {403},
  pages   = {97--117},
  year    = {2005},
  doi     = {10.1016/j.laa.2005.01.011}
}

@article{Barth1967,
  author  = {Barth, W. and Martin, R. S. and Wilkinson, J. H.},
  title   = {Calculation of the eigenvalues of a symmetric tridiagonal matrix by the method of bisection},
  journal = {Numer. Math.},
  volume  = {9},
  pages   = {386--393},
  year    = {1967},
  doi     = {10.1007/BF02162154}
}

@article{Domany1983,
  author  = {Domany, E. and Alexander, S. and Bensimon, D. and Kadanoff, L. P.},
  title   = {Solutions to the {Schr\"odinger} equation on some fractal lattices},
  journal = {Phys. Rev. B},
  volume  = {28},
  pages   = {3110--3123},
  year    = {1983},
  doi     = {10.1103/PhysRevB.28.3110}
}

@article{Bollt2005,
  author  = {Bollt, Erik M. and ben-Avraham, Daniel},
  title   = {What is special about diffusion on scale-free nets?},
  journal = {New J. Phys.},
  volume  = {7},
  pages   = {26},
  year    = {2005},
  doi     = {10.1088/1367-2630/7/1/026}
}

@article{BerkerOstlund1979,
  author  = {Berker, A. N. and Ostlund, S.},
  title   = {Renormalisation-group calculations of finite systems: order parameter and specific heat for epitaxial ordering},
  journal = {J. Phys. C},
  volume  = {12},
  pages   = {4961--4975},
  year    = {1979},
  doi     = {10.1088/0022-3719/12/22/035}
}

@book{GolubVanLoan,
  author    = {Golub, Gene H. and Van Loan, Charles F.},
  title     = {Matrix Computations},
  edition   = {4th},
  publisher = {Johns Hopkins University Press},
  address   = {Baltimore},
  year      = {2013}
}

@article{Dean1972,
  author  = {Dean, P.},
  title   = {The vibrational properties of disordered systems: numerical studies},
  journal = {Rev. Mod. Phys.},
  volume  = {44},
  pages   = {127--168},
  year    = {1972},
  doi     = {10.1103/RevModPhys.44.127}
}

@article{Montroll1965,
  author  = {Montroll, Elliott W. and Weiss, George H.},
  title   = {Random walks on lattices. {II}},
  journal = {J. Math. Phys.},
  volume  = {6},
  pages   = {167--181},
  year    = {1965},
  doi     = {10.1063/1.1704269}
}

@article{Arenas2006,
  author  = {Arenas, Alex and D\'{\i}az-Guilera, Albert and P\'erez-Vicente, Conrad J.},
  title   = {Synchronization reveals topological scales in complex networks},
  journal = {Phys. Rev. Lett.},
  volume  = {96},
  pages   = {114102},
  year    = {2006},
  doi     = {10.1103/PhysRevLett.96.114102}
}

@article{RadicchiArenas2013,
  author  = {Radicchi, Filippo and Arenas, Alex},
  title   = {Abrupt transition in the structural formation of interconnected networks},
  journal = {Nat. Phys.},
  volume  = {9},
  pages   = {717--720},
  year    = {2013},
  doi     = {10.1038/nphys2761}
}

@article{Gomez2013,
  author  = {G\'omez, S. and D\'{\i}az-Guilera, A. and G\'omez-Garde\~nes, J. and P\'erez-Vicente, C. J. and Moreno, Y. and Arenas, A.},
  title   = {Diffusion dynamics on multiplex networks},
  journal = {Phys. Rev. Lett.},
  volume  = {110},
  pages   = {028701},
  year    = {2013},
  doi     = {10.1103/PhysRevLett.110.028701}
}

\end{document}